\documentclass{article}
\usepackage{graphicx} 
\usepackage{amsmath}
\usepackage{amssymb}
\usepackage[colorlinks, citecolor=blue, urlcolor=blue,linkcolor=black]{hyperref}
\usepackage{natbib}
\usepackage{float}
\usepackage{authblk}
\usepackage{algorithm}
\usepackage{algpseudocode}
\usepackage{booktabs}
\usepackage{parskip}
\usepackage[top=2cm, bottom=2cm, left=2.5cm, right=2.5cm]{geometry} 
\title{Percolation Critical Probability of Aperiodic Smith Hat tile(1, $\sqrt3$)\thanks{Code available at: \url{https://github.com/aaryashBharadwaj/Aperiodic-Monotile-Percolation}}}

\newcommand{\equalcontrib}{\textsuperscript{$\dagger$}}
\author[1]{Haitao Gao\equalcontrib\thanks{$^\dagger$These authors contributed equally to this work. Corresponding author: \texttt{haitao.gao@unsw.edu.au}}}
\author[2]{Aaryash Bharadwaj\equalcontrib}
\affil[1]{School of Computer Science and Engineering, University of New South Wales, Sydney, NSW 2052, Australia}
\affil[2]{School of Mathematics and Statistics, The University of Sydney, Sydney, NSW 2006, Australia}
\date{December 2025}

\begin{document}

\maketitle

{\centering \section*{Abstract}}

The Smith Hat tile is the first known aperiodic monotile, having been discovered in 2023. The simple structure, constructed using only 8 kites, is unique and well motivated for analysis within percolation theory. The primary goal of this paper is to discover the critical threshold $p_c$ in both site and bond Bernoulli structures using Monte Carlo simulation for the Smith hat tile(1,$\sqrt3$). Our findings are site and bond values of $p_c^s = 0.822725 \pm 0.000044$ and $p_c^b = 0.798161 \pm 0.000044$  for edge percolation and $0.544247 \pm 0.000101$ for site percolation on the dual graph. 

\section*{Introduction}

A tiling of the plane is an arrangement of one or more shapes that covers the entire plane without
gaps or overlaps. For a tiling to be periodic, it must possess translational symmetry. This is satisfied by most traditional tilings, which come under the Wallpaper groups, a mathematical classification of a two-dimensional repetitive pattern based on the symmetries \citep{fedorov1891simmetrija, polya1924xii}. Aperiodic tilings have no such translational symmetry and are thus geometrically unique. 

The search for a single geometric shape that can tile the plane aperiodically, the Einstein problem, so-called because it means 
``one stone'' stood as a major open challenge in geometry for over half a century. Whilst famous examples like the Penrose tiling \citep{penrose1979pentaplexity} achieve aperiodicity, this relies on using at least two distinct shapes. 

The discovery of the Smith Hat tile in 2023 by \cite{smith2024aperiodic} provided the first solution to this problem, using a single 13-sided aperiodic monotile, referred to as the Smith Hat tile, to break translational invariance. 

Such tilings, beyond their importance within geometry itself, serve as the mathematical foundation for understanding quasicrystals, materials that exhibit long-range order and diffraction patterns but lack the translational symmetry characteristic of traditional periodic crystals \citep{deguchi2015superconductivity,kamiya2018discovery,fuchs2016hofstadter}. 

The unique aperiodic geometry of the Smith Hat tile raises natural questions about how local connectivity constraints influence global phenomena, making percolation theory a particularly well-suited framework for its analysis.
 

Percolation theory is a mathematical framework in statistical physics and probability that models the emergence of large-scale connectivity in random media \citep{duminil2018sixty}. The theory examines the Bernoulli percolation process, in which a parameter $p$ determines whether an element of an infinite lattice is open or closed. Within this there are two primary models: bond percolation, where edges of a lattice are independently open with probability $p$ and otherwise closed, and site percolation where the vertices are independently open or closed with probability $p$. Elements that are open and adjacent form connected clusters.

The central question in both variants of Bernoulli percolation is to determine the critical probability
$p_c$: the threshold value such that, on an infinite tiling, an infinite cluster appears for all 
$p \ge p_c$ and does not exist for all $p < p_c$. At $p_c$, a unique infinite cluster emerges with fractal properties, exhibiting scaling behaviours characterised by the critical exponent $\nu = 4/3$ \citep{nienhuis1987coulomb} in two dimensions under a universal  hypothesis \citep{langlands1992universality}.

This has wide-reaching implications for a myriad of structures, where local structure in a large system leads to a global phase transition.

For infinite regular lattices in two dimensions, many percolation thresholds $p_c$ are known analytically or to high precision, forming the foundation of percolation theory. For the square-lattice (coordination number $z = 4$), the bond percolation threshold is exactly $p_c^b=0.5$, a result established by self-duality and rigorous proof \citep{kesten74critical,nolin2008critical}. The site percolation threshold on the square lattice is $p_c^s\approx0.592746$, known from high-precision Monte Carlo simulations \citep{derrida1985corrections,newman2000efficient,lee2008pseudo, jacobsen2014high, yang2024comment}. 

For triangular lattice ($z=6$), which is the planar dual of the hexagonal lattice, the site threshold is exactly $p_c^s=1/2$ whilst the bond threshold is known to be exactly $p_c^b=2\sin(\pi/18)\approx0.347296$ \citep{sykes1964exact}, notably the lowest among 2D lattices with triangular connectivity. 

The honeycomb (hexagonal) lattice ($z=3$), dual to the triangular, has $p_c^s\approx0.697041$ \citep{djordjevic1982site,ziff2009universal,jacobsen2014high} and $p_c^b\approx0.652704$ (the latter equals $1-2\sin(\pi/18)$ by duality) \citep{sykes1964exact}. Other Archimedean lattices show similar thresholds in the range ~$0.5$–$0.8$ for site percolation, depending on coordination and lattice geometry. In general, higher coordination lowers the threshold, but lattice topology complicates this and can create variation.

Since percolation is fundamentally concerned with the nodes and edges of a graph, there are several ways to define a graph structure, depending on what is designated as a node and an edge. Both bond and site percolation can then be performed on any such graph, as they depend only on the graph's structure rather than the underlying tiling itself.

What we refer to as \textit{edge percolation} treats each vertex of the tiling as a node and each edge of the tiling as an edge. This is the primary method used to obtain the results shown above. The second approach, which we refer to as \textit{tile percolation} as well as the 'dual graph' in the abstract and codebase, takes each tile as a node with an edge existing between two nodes if and only if the corresponding tiles are adjacent. 

For periodic tiling, the distinction is rarely emphasised since tile percolation is simply the percolation of the dual graph. For bond percolation on a tiling, the critical thresholds for tile percolation and edge percolation sum to 1 which means that a result for one immediately implies the other \citep{beffara2006percolation}. The same methodology is applied to both throughout, the only difference being a distinct graph builder that constructs the graph by treating tiles, rather than vertices, as nodes.

The extension of percolation theory from periodic lattices to aperiodic structures, such as the Smith Hat tile, introduces unique mathematical challenges primarily due to the loss of translational symmetry. In a standard periodic lattice such as $\mathbb{Z^2}$, the probability of percolation is identical for all sites due to translation invariance; however, in aperiodic tilings, the vertices have varying local environments (patches). This means that the probability that a specific vertex belongs to an infinite cluster is vertex dependent. 

\cite{Lu1987QuasilatticePercolation} initiated the study of percolation on non-periodic structures through their work on Penrose quasilattices, motivated by the discovery of quasicrystals and their non-periodic long-range order. Using computer simulations on 50{,}000 sites, they estimated the bond-percolation threshold as $p_c^{\text{bond}} \approx 0.483 \pm 0.005$ and concluded that it is consistent with the universality class for percolation on periodic two-dimensional lattices. Around the same time, in 1989, \cite{Sakamoto1989PercolationThresholds} employed a Monte Carlo approach and proposed a slightly different numerical estimate of $p_c^{\text{bond}} \approx 0.477 \pm 0.002$, based on simulations of 90{,}000 sites. \cite{hof1998percolation} further showed analytically that aperiodic tilings have a consistent result within combined error bars, motivating the computational analysis within our work on the Smith Hat monotile. 

More recently, in 2024, the Ising model was applied to the aperiodic \emph{hat} tile that is the subject of our work, and researchers found a critical temperature of $T_c/J \approx2.405  \pm0.0005$ using a similar Monte-Carlo method \citep{Okabe2024SmithHatIsing}. This research is relevant as both critical temperature and critical threshold identify a point of phase transition where universal critical phenomena such as scaling and power-law behaviour emerge.

However, it is essential to distinguish these thermal phase transitions from the geometric phase transitions characterised by percolation. While the Ising model focuses on the temperature-dependent correlation and alignment of spins, percolation focuses on the fundamental connectivity of the underlying lattice. Because thresholds $T_c$ and the percolation critical probability $p_c$ are not derivable from one another, the specific geometric arrangement of the Smith Hat tile necessitates a dedicated investigation into its connectivity limits. 

Despite the growing body of work on aperiodic monotiles, the site and bond percolation thresholds for the Smith Hat tile's geometry have not yet been established in the literature. This research aims to address this gap by performing extensive Monte Carlo simulations on the Smith-kite lattice. Our objective is to utilise Finite-Size Scaling (FSS) analysis, specifically leveraging the $L^{-1/\nu}$ scaling relation to accurately determine the infinite system threshold $p_c^s$ and $p_c^b$. 

Our findings indicate critical probabilities of $p_c^s \approx 0.822725$ for the site percolation and $p_c^b \approx 0.798161$ for bond percolation. These values are higher than those observed among studied 2D lattices, suggesting that the unique ``hat" geometry possesses distinct connectivity constraints compared to more common periodic or aperiodic structures. By providing these first estimates, this study establishes a benchmark for the geometric phase behaviour of the first discovered aperiodic monotile. 

\section*{Model and Numerical Method}

\subsection*{Lattice}
\begin{figure}[H]
    \centering
    \includegraphics[width=0.4\textwidth]{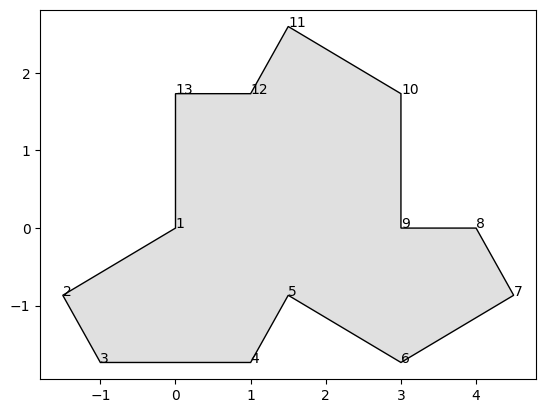}
    \caption{Hat unit}
    \label{fig: hat unit}
\end{figure}

\begin{figure}[H]
    \centering
    \includegraphics[width=0.7\textwidth]{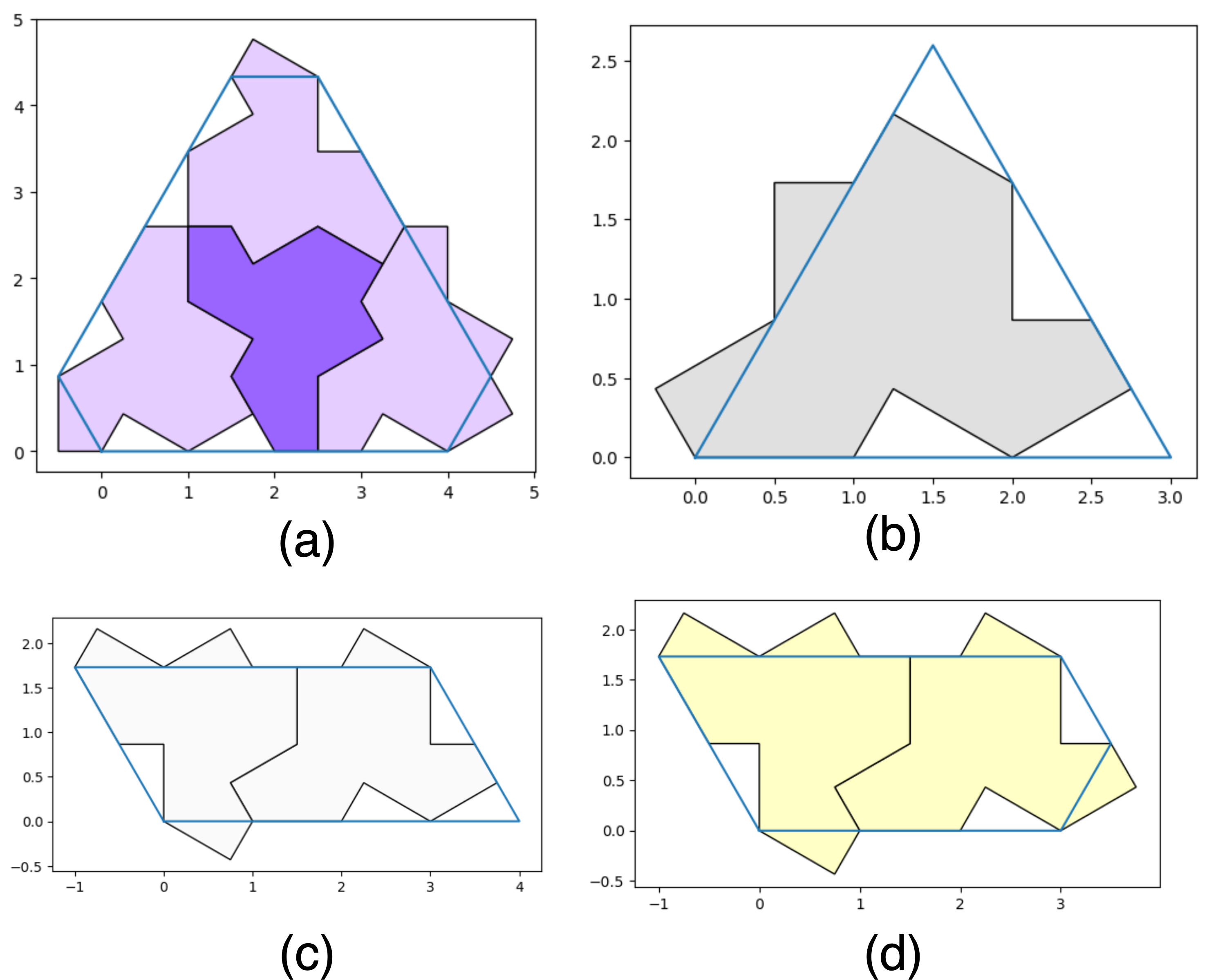}
    \caption{Metatile in Smith hat tiling: (a) pattern $H$, (b) pattern $T$, (c) pattern $P$, (d) pattern $F$}
    \label{fig: Metatile}
\end{figure}

The Smith hat tiling is composed of one single hat prototile as shown in Figure~\ref{fig: hat unit}. Tiles can be grouped into four types of clusters, each containing between one and four hats. The boundaries of these clusters are then abstracted to produce four \textbf{metatiles}, labelled $H$ (an irregular hexagon), $T$ (an equilateral triangle), $P$ (a parallelogram), and $F$ (a pentagonal shape), as shown in Figure~\ref{fig: Metatile} \citep{smith2024aperiodic}.

These four metatiles inherit matching rules from the boundaries of the hats they contain, constraining how they may be placed adjacent to one another. Specifically, the $H$ metatile corresponds to a cluster of four hats (one reflected and three unreflected), the $T$ metatile to a single isolated hat, the $P$ metatile to a pair of hats forming a parallelogram-shaped cluster, and the $F$ metatile to a pair of hats associated with one leg of a three-armed propeller structure called a \textit{triskelion}.

\begin{figure}[H]
    \centering
    \includegraphics[width=0.8\textwidth]{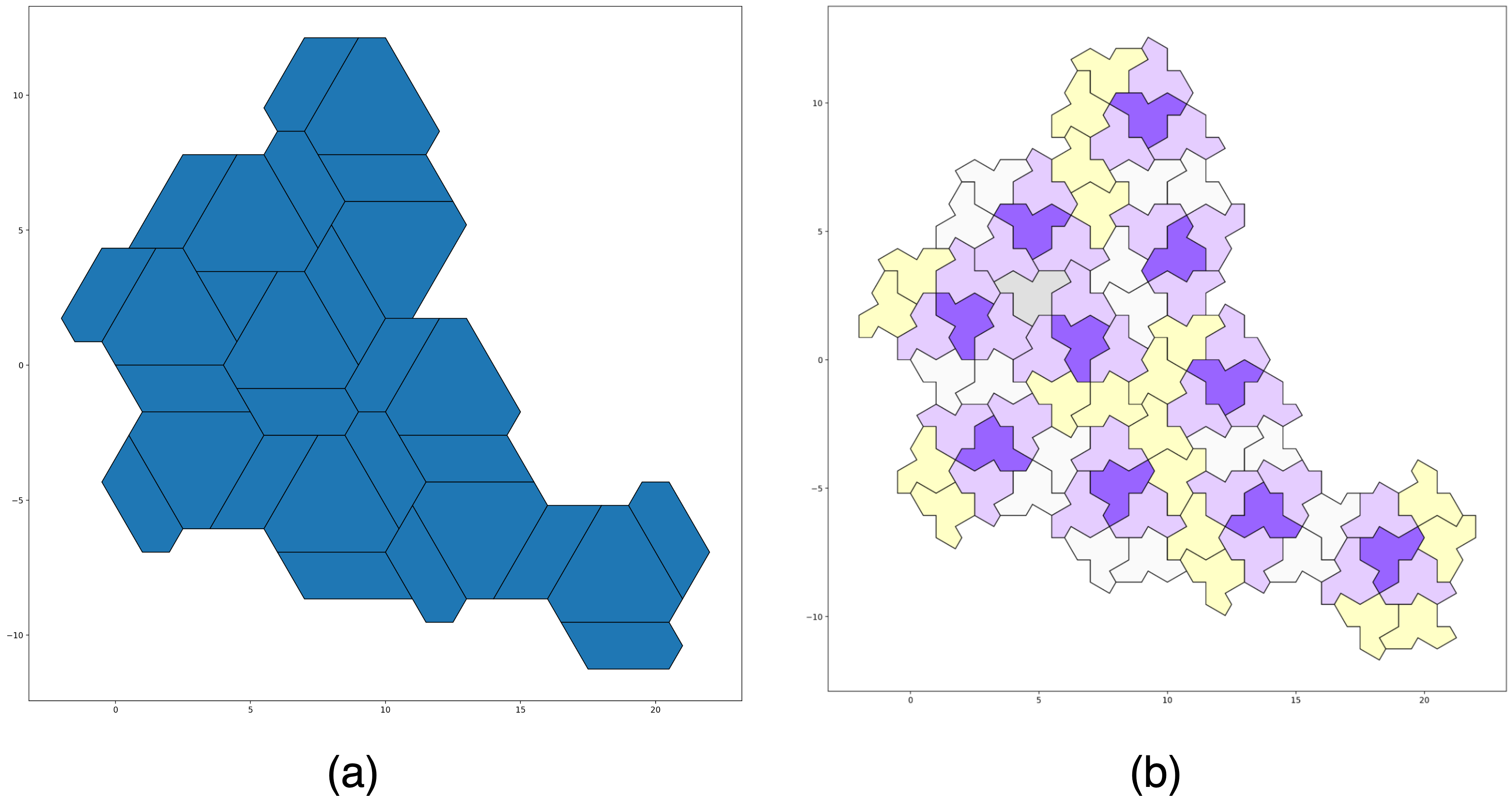}
    \caption{Hat tile: (a) The patch composed by metatiles. (b) The metatiles dissected into multiple copies of the prototile.}
    \label{fig: hattile}
\end{figure}

The Smith hat tiling is a substitution tiling, meaning it can be generated through an iterative inflate-and-subdivide process applied to the metatiles. At each iteration, a patch of metatiles is assembled in a fixed configuration (as specified by the substitution rules in \citep{smith2024aperiodic}), and key vertices within this patch define the outlines of larger supertiles. These supertiles are not geometrically similar to the original metatiles (with the exception of $T$), but they are combinatorially equivalent: they have the same connectivity and obey the same matching rules. This means the substitution can be applied again to the supertiles, producing level-2 supertiles, and so on indefinitely. At each level, the corresponding hat clusters are carried along, so that iterating the process generates arbitrarily large patches of the hat tiling. Figure~\ref{fig: hattile} shows a hat tiling patch generated by one iteration of this substitution rule.

The tiling generation follows the seminal paper \citep{smith2024aperiodic}, whose original JavaScript implementation was transcribed and adapted to Python for this work. Further details on the construction are found in that paper.

\subsection*{Monte Carlo Simulation}
Monte Carlo simulation is a common method to estimate the critical probability $p_c$ for geometries that are not possible to calculate the theoretical value of. For percolation theory, a finite subset of the infinite lattice is generated to estimate the critical probability for an infinite cluster. This is done by checking if a connected cluster exists from one boundary to another of the finite subset. 

A cluster could percolate rightward which means some element on the left boundary is in the same cluster as one on the right, or downward which means the same regarding the upper and lower boundaries. We therefore define the following estimators:

\begin{itemize}
    \item $p_L^R(T)$: the sample probability that we find a rightward percolating cluster in the framed square with length $L$ with $T$ independent simulations.
    \item $p_L^D(T)$: the sample probability that we find a downward percolating cluster in the framed square with length $L$ with $T$ independent simulations.
    \item $p_L^I(T)$: the sample probability that we find a cluster percolating both rightward and downward in the framed square with length $L$ with $T$ independent simulations.
    \item $p_L^U(T)$: the sample probability that we find a cluster percolating either rightward or downward in the framed square with length $L$ with $T$ independent simulations.
    \item $p_L^A(T) = \frac{1}{2}[p_L^R(T) + p_L^D(T)] = \frac{1}{2}[p_L^I(T) + p_L^U(T)]$, assuming that the tiling is isotropic. This is suggested  by \cite{RIEGER2024104988} which states the Smith Hat tiling behaves as an isotropic elastic continuum at large scales and suggesting the absence of any preferred directional bias in the tiling's macroscopic structure.
\end{itemize}
The notation $R,D, I, U$ and $A$ respectively represent rightward, downward, intersection, union and average. It follows directly:
\begin{equation}
    p_L^I(T) \le p_L^A(T) \le p_L^U(T)
\end{equation}
And thus,  $p_L^I(T)$ and $p_L^U(T)$ bound the true threshold and convergence of the two narrows the bound of the final value. 

The estimator $p_L^X = \bar{p}$ is taken as the sample mean:
\begin{equation}
    \bar{p} = \frac{p_1 + p_2 + \cdots + p_T}{T}
\end{equation}
where $p_i = \frac{M_i}{N}$ with $M_i$ denoting the number of sites opened at the moment percolation occurs in trial $i$, and $N$ is the total number of sites in the lattice. The standard deviation is given by
\begin{equation}
    s^2 = \frac{(p_1 - \bar{p})^2 + (p_2 - \bar{p})^2 + \cdots + (p_T - \bar{p})^2}{T - 1}
\end{equation}

By the Central Limit Theorem, for sufficiently large $T$, $\bar{p}$ is approximately normally distributed, giving the following 95\% confidence interval for the percolation threshold:
\begin{equation}
    [\bar{p} - \frac{1.96s}{\sqrt{T}}, \bar{p} + \frac{1.96s}{\sqrt{T}}]
\end{equation}

The framed squares were centred on the coordinate (200, -100) to maximise the size of the frame we could sample from, as shown below. Having it centred on a point outside the patch would leave part of the sampled region empty and thus harm the results.

\begin{figure}[H]
    \centering
    \includegraphics[width=\textwidth]{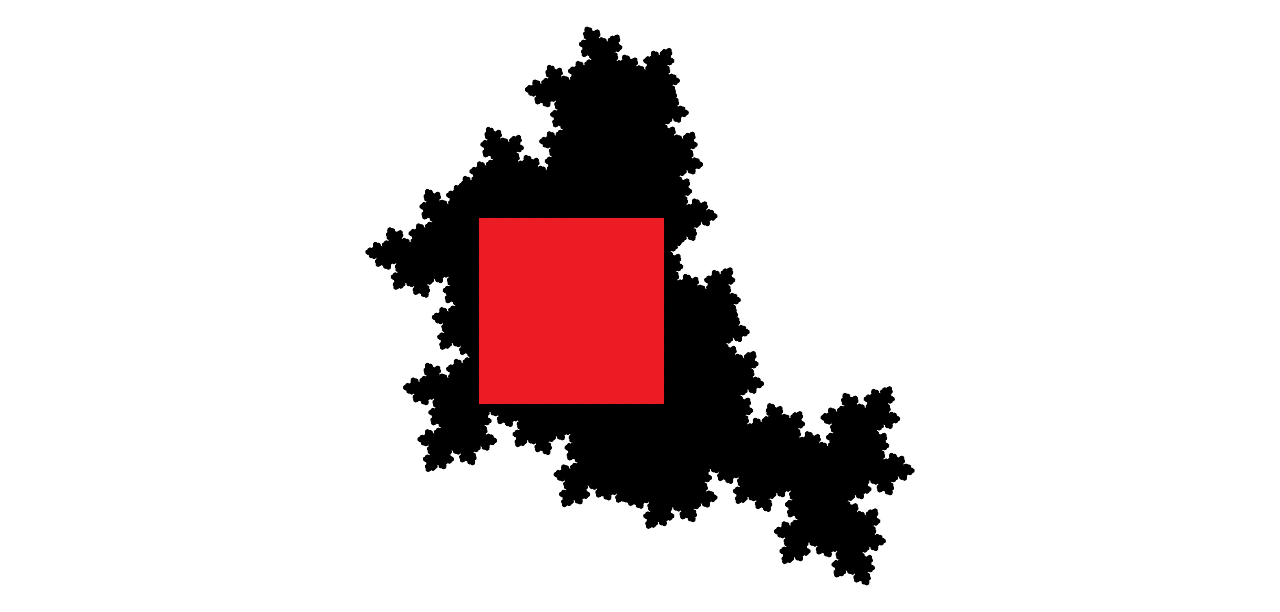}
    \caption{Patch 5: The Red Square is the L = 400 centred on the point (200, -100)}
    \label{fig: frames}
\end{figure}

\section*{Results}
We generate a patch with five recursions and frame the square within the patch by length $L_{min} = 10$ to $L_{max} = 400$ with step 10 although the code provided allows independent testing at desired precision as shown in Figure~\ref{fig: frames}. We conduct 1000 independent experiments for each length. This combination was a compromise between computational cost and accuracy but various combinations were tested throughout the research process. 

\begin{figure}[H]
    \centering
    \includegraphics[width=\textwidth]{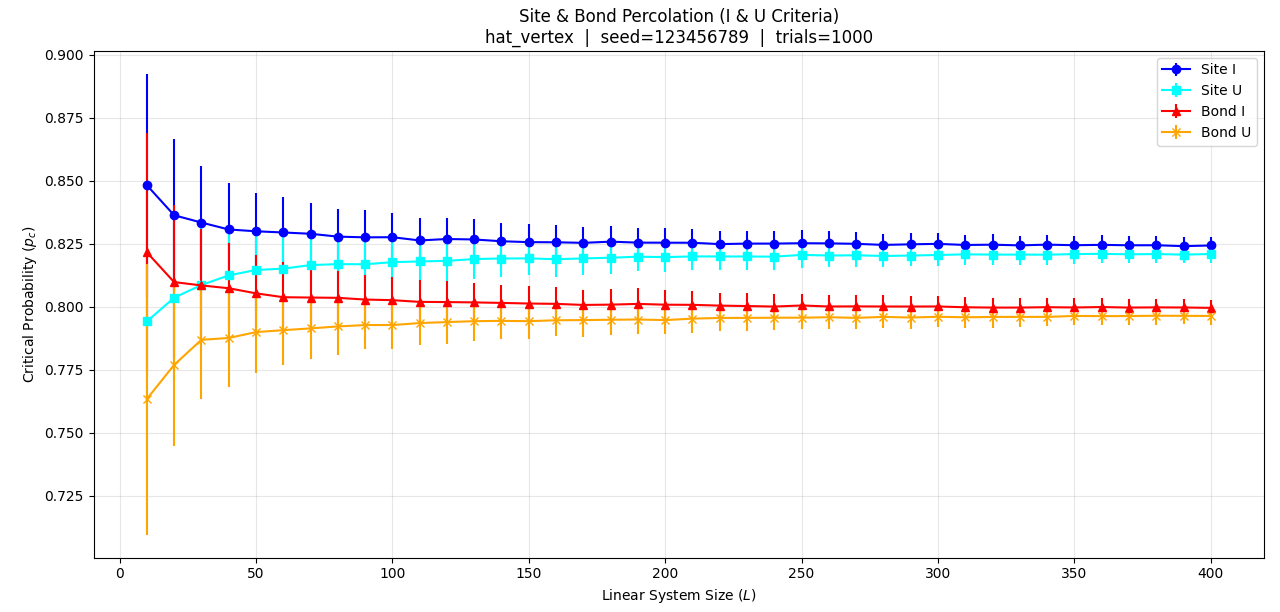}
    \caption{Monte Carlo simulation: Site and bond percolation critical probability mean $\bar{p_c}$ and 95\% confidence interval}
    \label{fig: mean and std}
\end{figure}

Figure ~\ref{fig: mean and std} shows the mean critical probability within a 95\% confidence interval. It follows the statistical property that when the system size $L$ increases, the 95\% confidence interval shrinks. 

Since our method of measuring was weighted-least squares, with a small sample size and an unknown population deviation, the t-distribution was the more robust and natural statistical measure as opposed to the normal. 

In order to estimate the critical probability of an infinite system, we extrapolate the $p_c^X(L)$ to $L \rightarrow \infty$ due to the scaling law,
\begin{equation}
    p_c^X(L) = p_c + A^X L^{-1/\nu}
\end{equation}
where $p_c^X(L)$ is the finite-size estimator, $p_c$ is the true critical probability when $L \rightarrow \infty$, $A^X$ is a scaling amplitude and $L$ is the system size. Previous percolation research in Penrose tiling \citep{hof1998percolation} and Ising model about Smith Hat tile \citep{Okabe2024SmithHatIsing} assume that they share the same exponent value $\nu = 4/3$. Therefore, we accept the universal hypothesis and set the same value in our experiments. This leads to strong numerical convergence and linear plots with Figures 5 and 6, however it is not strictly proven.

To estimate the uncertainty of the extrapolated $p_c$, we reformulate the scaling law as a linear regression by setting $x_i = L_i^{-3/4}$, so that
\begin{equation}
    p_c^X(L_i) = p_c + A^X x_i,
\end{equation}
and apply weighted least squares (WLS), where each observation is weighted by $w_i = 1/\sigma_i^2$ with $\sigma_i = s_i/\sqrt{T_i}$ being the standard error of the mean obtained from the Central Limit Theorem, $s_i$ the sample standard deviation and $T_i$ the number of trials at system size $L_i$. Larger systems yield smaller $\sigma_i$ and therefore receive higher weight, reflecting their greater statistical precision. The WLS estimator of the intercept $p_c$ and its variance are given by
\begin{equation}
    \hat{p}_c = \frac{\sum w_i x_i^2 \sum w_i p_c^X(L_i) - \sum w_i x_i \sum w_i x_i p_c^X(L_i)}{\sum w_i \sum w_i x_i^2 - \left(\sum w_i x_i\right)^2},
\end{equation}
\begin{equation}
    \mathrm{Var}(\hat{p}_c) = \frac{\sum w_i x_i^2}{\sum w_i \sum w_i x_i^2 - \left(\sum w_i x_i\right)^2}.
\end{equation}
Note that $\mathrm{Var}(\hat{p}_c)$ depends only on the weights $w_i$ and the predictors $x_i$, not on the response values $p_c^X(L_i)$, since the uncertainty of the intercept is fully determined by the design of the experiment. A $95\%$ confidence interval for the true critical probability is then
\begin{equation}
    \hat{p}_c \pm t^*_{n-2} \cdot \sqrt{\mathrm{Var}(\hat{p}_c)},
\label{eq: extra-ci}
\end{equation}
where $t^*_{n-2}$ is the critical value of the $t$-distribution with $n-2$ degrees of freedom and $n$ is the number of system sizes used in the extrapolation.

\begin{figure}[H]
    \centering
    \includegraphics[width=\textwidth]{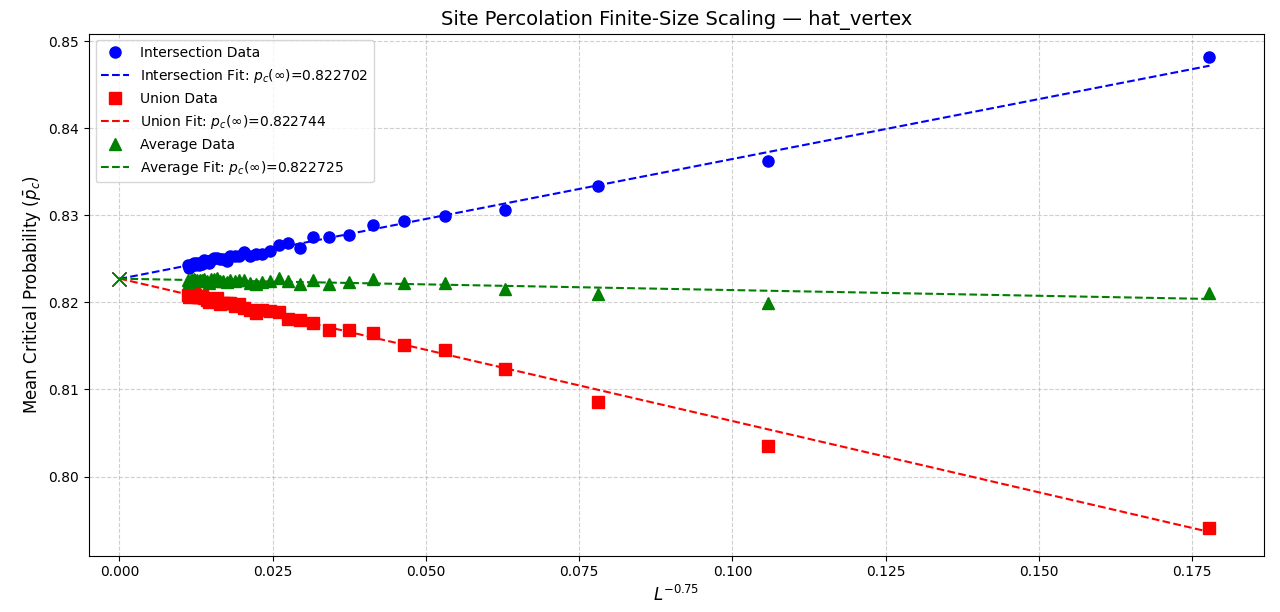}
    \caption{Site Percolation}
    \label{fig: site}
\end{figure}

\begin{figure}[H]
    \centering
    \includegraphics[width=\textwidth]{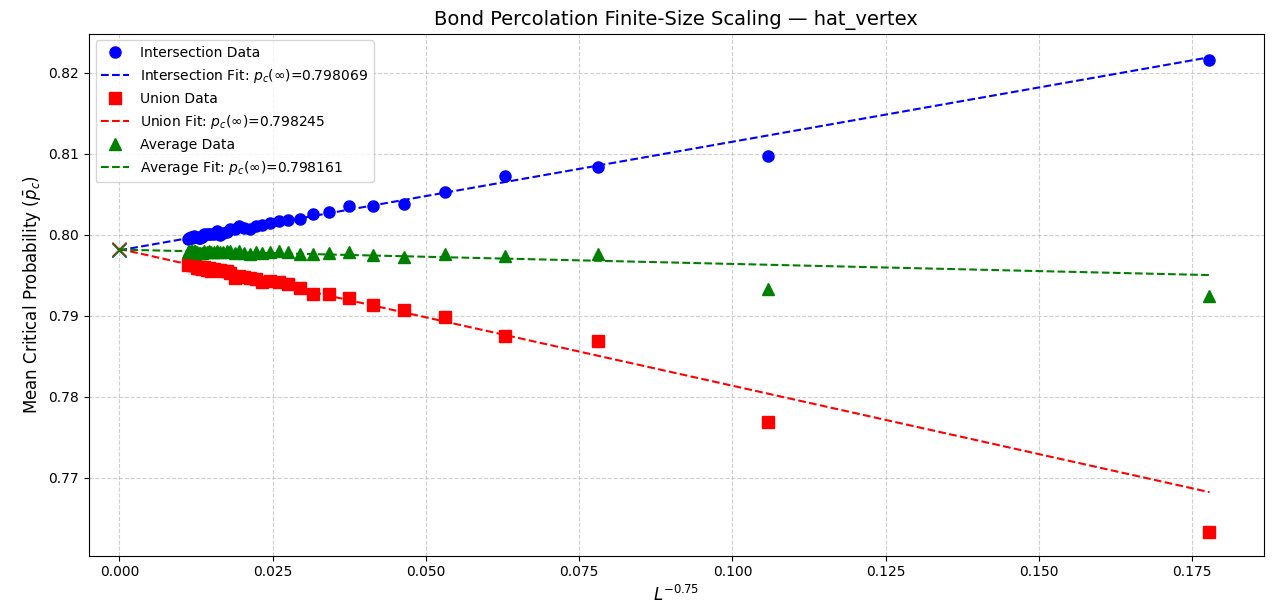}
    \caption{Bond Percolation}
    \label{fig: bond}
\end{figure}

Figure~\ref{fig: site} and Figure~\ref{fig: bond} show the site and bond percolation respectively. We can see that all intersection, union and average critical probabilities converge to the same value, that is the critical probability in the infinite system.

\begin{table}[H]
\centering
\caption{Monte Carlo simulation results: Site and Bond percolation probabilities for different system sizes $L$.}
\label{tab:percolation_results}
\begin{tabular}{c c c}
\hline
\textbf{System Size $L$} & \textbf{Site $p_c^s$ (Mean)} & \textbf{Bond $p_c^b$ (Mean)} \\
\hline
10  & 0.848159 & 0.821553 \\
20  & 0.836246 & 0.809724 \\
30  & 0.833343 & 0.808408 \\
40  & 0.830604 & 0.807288 \\
50  & 0.829879 & 0.805266 \\
60  & 0.829406 & 0.803730 \\
70  & 0.828893 & 0.803594 \\
80  & 0.827758 & 0.803480 \\
90  & 0.827475 & 0.802802 \\
100 & 0.827531 & 0.802552 \\
110 & 0.826245 & 0.801875 \\
120 & 0.826814 & 0.801800 \\
130 & 0.826633 & 0.801668 \\
140 & 0.825923 & 0.801467 \\
150 & 0.825577 & 0.801223 \\
160 & 0.825520 & 0.801088 \\
170 & 0.825290 & 0.800661 \\
180 & 0.825767 & 0.800785 \\
190 & 0.825383 & 0.801068 \\
200 & 0.825353 & 0.800763 \\
210 & 0.825329 & 0.800698 \\
220 & 0.824792 & 0.800390 \\
230 & 0.824993 & 0.800223 \\
240 & 0.824999 & 0.800019 \\
250 & 0.825144 & 0.800445 \\
260 & 0.825081 & 0.800047 \\
270 & 0.824910 & 0.800091 \\
280 & 0.824474 & 0.800047 \\
290 & 0.824710 & 0.800026 \\
300 & 0.824893 & 0.800059 \\
310 & 0.824445 & 0.799737 \\
320 & 0.824562 & 0.799616 \\
330 & 0.824289 & 0.799619 \\
340 & 0.824579 & 0.799774 \\
350 & 0.824350 & 0.799654 \\
360 & 0.824497 & 0.799827 \\
370 & 0.824323 & 0.799619 \\
380 & 0.824348 & 0.799696 \\
390 & 0.823983 & 0.799647 \\
400 & 0.824254 & 0.799503 \\
$\infty$ & 0.822725 & 0.798161 \\
\hline
\end{tabular}
\end{table}

The table shows the values plotted and the extrapolated value for $L$ approaching infinity derived from the convergence of the intersection and union percolation extrapolations shown in the graph. This allows direct verification against independent implementations with the same random seed of $123456789$.


Based on Eq~\ref{eq: extra-ci}, we estimate the uncertainty of extrapolated site and bond percolation critical threshold. When the system size $L \rightarrow \infty$, $p_c^s = 0.822725$ with $\text{95\%  CI = [0.822636, 0.822815]}$ and $p_c^b = 0.798161$ with $\text{95\% CI = [0.798073, 0.798250]}$.

For tile percolation, since bond tile and bond edge thresholds sum to 1 by duality, only site percolation was simulated. Using the same methodology as the edge percolation, the results are as follows: 

\begin{figure}[H]
    \centering
    \includegraphics[width=\textwidth]{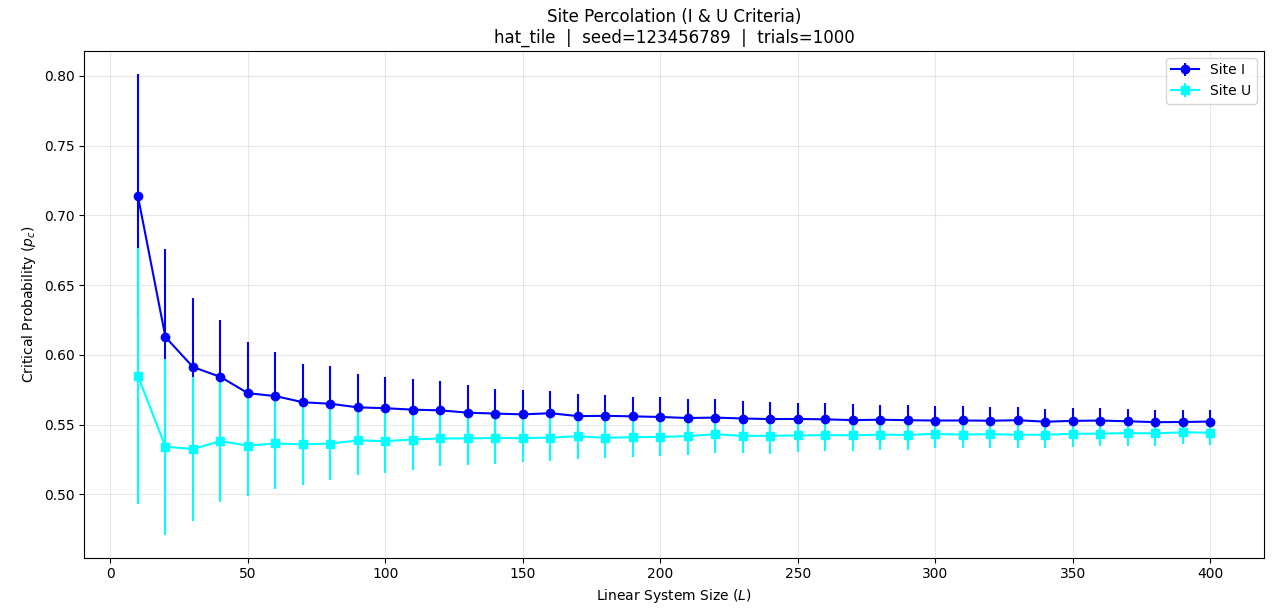}
    \caption{Monte Carlo simulation: Site percolation critical probability mean for the tile percolation $\bar{p_c}$ and 95\% confidence interval}
    \label{fig: mean and std tile}
\end{figure}

\begin{figure}[H]
    \centering
    \includegraphics[width=\textwidth]{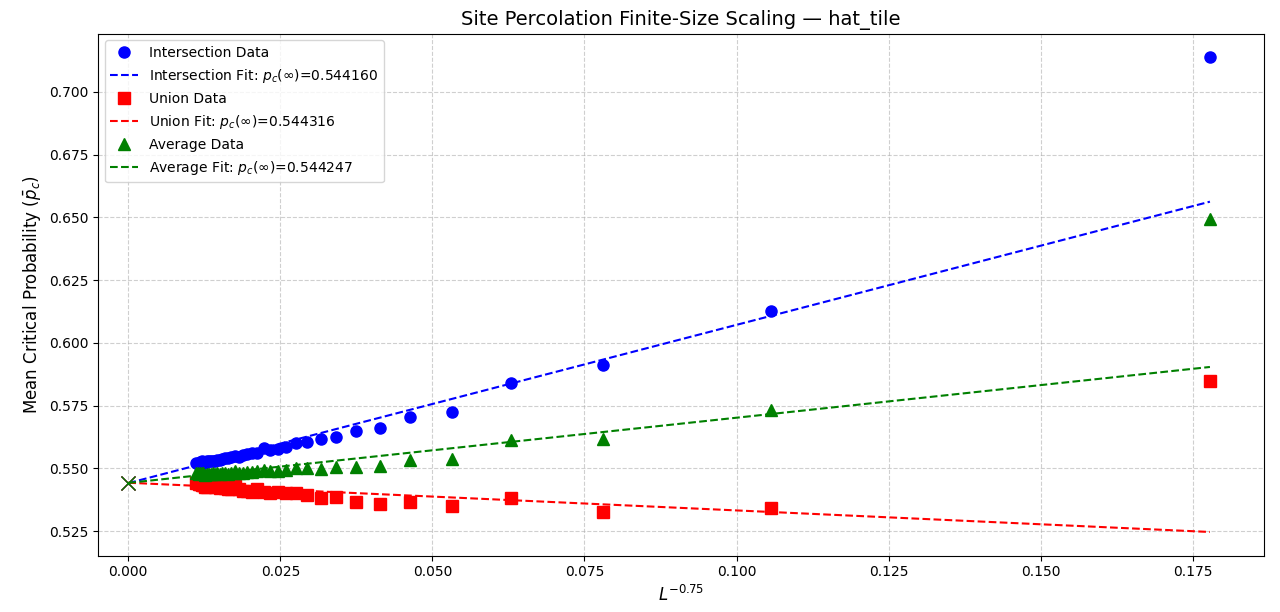}
    \caption{Site Percolation of Tile}
    \label{fig: Site_Tile}
\end{figure}




When the system size for the tile percolation $L \rightarrow \infty$, $p_c^s = 0.544247$ with $\text{95\% CI = [0.544044, 0.544450]}$

In Summary:
\begin{table}[H]
\centering
\begin{tabular}{l c c c c}
\toprule
 & \textbf{Site} $p_c^s$ & \textbf{95\% CI} & \textbf{Bond} $p_c^b$ & \textbf{95\% CI} \\
\midrule
Edge & $0.822725$ & $[0.822636,\ 0.822815]$ & $0.798161$ & $[0.798073,\ 0.798250]$ \\
\midrule
Tile & $0.544247$ & $[0.544044,\ 0.544450]$ & $0.201839$ & $[0.201750,\ 0.201927]\ {\small(\text{by duality})}$ \\
\bottomrule
\end{tabular}
\end{table}

\section*{Summary and Discussion}

In this work, we have determined the site and bond percolation critical probabilities for the Smith hat tile, finding $0.822725 \pm 0.000044$ for site percolation and $0.798161 \pm 0.000044$ for bond percolation. These results are notably high compared to other studied 2D lattices. This can be understood through the coordination number.  

Percolation thresholds tend to be inversely proportional to coordination number \citep{Xun_2020}, since higher connectivity reduces the occupation probability required to form a spanning cluster. Since the Smith Hat tile is aperiodic, it lacks a fixed coordination number; however, using the average coordination number of approximately 2.31 reported by \cite{PhysRevLett.132.086402}, the high critical probabilities are consistent with this low average connectivity. This is also seen geometrically where the tile has places with 2 bonds in parallel. The tile percolation's site percolation results of $0.544247 \pm 0.000101$ fit within the fairly standard threshold range of roughly $0.4 -0.7$. 

These results establish the first percolation benchmarks for the Smith Hat tile geometry and provide a foundation for future analytical work on aperiodic monotiles. A natural direction for future research would be to investigate whether the critical exponents confirm the expected 2D universality class, and to extend this analysis to other members of the Hat tile family.

The physical implications of these results are significant. In the context of electrical networks subject to random disruptions, the high percolation threshold implies that a greater fraction of components must fail before global connectivity is lost. This has potential relevance for the design of fault-tolerant networks and for understanding transport properties in quasicrystalline materials, where the Smith Hat tiling serves as a mathematical model. 

\section*{Limitations}

The primary computational limitation of this work is precision. While our results achieve confidence intervals on the order of $\pm 0.00005$, greater precision is attainable through parallelisation, faster union-find implementations, increased trial counts and a compiled language. The current implementation is sequential and was constrained by available compute time.

The results rely on two major assumptions. First, that the Smith Hat tile is percolation-isotropic which justifies the average estimator $p_L^A$. Secondly, we assume the critical exponent $\nu = \frac{4}{3}$, though this is true for all percolation in 2-dimensions where there is no long-range connections which is expected to be the case here. Moreover numerical convergence heavily suggests this and precedent from the Penrose tiling and Ising model work further reinforce this, it has not been analytically proven for this geometry.

Lastly, whilst the results generalise to many of the other Hat tiles beyond the $(1, \sqrt3)$ that is our primary subject, others may have unique results worth exploring independently.

\section*{Acknowledgement}
The authors would like to thank Manav Garg for proposing the study of the percolation critical value on Smith Hat tile, John Zhang for his implementation of the Penrose tiling generation, Vladimir Pak for his research and writing, Dr.~Simon Mackenzie for his mentorship and valuable guidance throughout this project and Prof.~Robert Ziff for his review and invaluable insight on this fascinating field. We also extend our gratitude to Prof.~Serge Gaspers for his support and instruction in the Extended Algorithm Design and Analysis course.

As an important note, AI language models were used as a tool in parts of the code implementations, grammar-checking and bug-testing although all outputs produced were verified and hand-checked rigorously. 

\appendix
\newpage
\section*{Appendix}
\subsection*{Algorithm}

\begin{algorithm}[H]
\caption{Hat Tiling Graph Construction (Edge Percolation)}
\label{alg:graph_builder}
\begin{algorithmic}[1]
    \State \textbf{Input:} Substitution patch $P$ at recursion level $\ell$
    \State \textbf{Output:} Unique node coordinates $V$, Adjacency list $Adj$
    \State Initialise empty node list $\mathbf{N} \leftarrow \emptyset$
    \State \textbf{Collect nodes:} Traverse patch hierarchy recursively to depth $\ell$
    \For{each leaf tile $t$ in patch}
        \For{each vertex $v$ of $t$}
            \State Append screen coordinates of $v$ to $\mathbf{N}$
        \EndFor
    \EndFor
    \State Build KDTree on $\mathbf{N}$; query all pairs within tolerance $\varepsilon = 10^{-5}$
    \State \textbf{Merge duplicates:} Initialise Union-Find $UF$ over $|\mathbf{N}|$ nodes
    \For{each coincident pair $(i, j)$}
        \State $\text{UNION}(UF, i, j)$
    \EndFor
    \State Map all nodes to canonical representatives; extract unique nodes $V$
    \State \textbf{Build edges:} Traverse patch hierarchy again to depth $\ell$
    \For{each leaf tile $t$ with vertices $v_0, v_1, \dots, v_{n-1}$}
        \For{$i = 0$ to $n-1$}
            \State $u \leftarrow$ canonical index of $v_i$, \quad $w \leftarrow$ canonical index of $v_{(i+1) \bmod n}$
            \If{$u \neq w$}
                \State Add edge $(u, w)$ to edge set
            \EndIf
        \EndFor
    \EndFor
    \State Build adjacency list $Adj$ from edge set
    \State \Return $V$, $Adj$
\end{algorithmic}
\end{algorithm}

\begin{algorithm}[H]
\caption{Hat Tiling Graph Construction (Tile Percolation)}
\label{alg:tile_graph_builder}
\begin{algorithmic}[1]
    \State \textbf{Input:} Substitution patch $P$ at recursion level $\ell$
    \State \textbf{Output:} Tile centroid coordinates $C$, Adjacency list $Adj$
    \State Initialise empty polygon list $\mathbf{T} \leftarrow \emptyset$
    \State \textbf{Collect tiles:} Traverse patch hierarchy recursively to depth $\ell$
    \For{each leaf tile $t$ in patch}
        \State Append vertex array of $t$ to $\mathbf{T}$
    \EndFor
    \State Compute centroid $c_i \leftarrow \text{mean}(T_i)$ for each tile $T_i \in \mathbf{T}$; store in $C$
    \State \textbf{Find shared edges:} Build flat array of all tile vertices with tile index labels
    \State Build KDTree on all vertices; query all coincident pairs within tolerance $\varepsilon = 10^{-5}$
    \For{each coincident vertex pair $(i, j)$ belonging to tiles $(t_i, t_j)$ with $t_i \neq t_j$}
        \State Increment shared vertex count for pair $(t_i, t_j)$
    \EndFor
    \State \textbf{Build adjacency:} Initialise empty adjacency list $Adj$ of size $|\mathbf{T}|$
    \For{each tile pair $(t_i, t_j)$ with shared vertex count $\geq 2$} \Comment{two shared vertices indicate a shared edge}
        \State Add $t_j$ to $Adj[t_i]$, \quad Add $t_i$ to $Adj[t_j]$
    \EndFor
    \State \Return $C$, $Adj$
\end{algorithmic}
\end{algorithm}

\begin{algorithm}[H]
\caption{Monte Carlo Simulation for Smith Hat tile Site Percolation (Intersection and Union)}
\label{alg:hat_percolation_site}
\begin{algorithmic}[1]
    \State \textbf{Input:} Set of nodes $V$ (size $N$), Adjacency list $Adj$,
    \Statex \hspace{4.2em} Boundary sets $S_{Top}, S_{Bot}, S_{Left}, S_{Right}$
    \State \textbf{Input:} Number of trials $T$, Percolation type $\mathcal{M} \in \{\text{Intersection}, \text{Union}\}$
    \State \textbf{Output:} Mean critical threshold $\bar{p}_c$, Standard deviation $\sigma$, 95\% CI
    \State Initialise results list $\mathbf{P} \leftarrow \emptyset$
    \For{$k = 1$ to $T$}
        \State Initialise site states $s \leftarrow \mathbf{0}$ of size $N$ \Comment{0 = closed, 1 = open}
        \State Initialise $UF_{TB}$ with virtual nodes $v_{top}, v_{bot}$
        \State Initialise $UF_{LR}$ with virtual nodes $v_{left}, v_{right}$
        \State $N_{open} \leftarrow 0$
        \State Generate a uniformly random permutation $\pi$ of $\{0, \dots, N-1\}$
        \For{each site $i$ in order $\pi$}
            \State $s[i] \leftarrow 1$, \quad $N_{open} \leftarrow N_{open} + 1$
            \Comment{Opening site $i$ activates its boundary and neighbour connections}
            \If{$i \in S_{Top}$} $\text{UNION}(UF_{TB},\, v_{top},\, i)$ \EndIf
            \If{$i \in S_{Bot}$} $\text{UNION}(UF_{TB},\, v_{bot},\, i)$ \EndIf
            \If{$i \in S_{Left}$} $\text{UNION}(UF_{LR},\, v_{left},\, i)$ \EndIf
            \If{$i \in S_{Right}$} $\text{UNION}(UF_{LR},\, v_{right},\, i)$ \EndIf
            \For{each neighbour $j \in Adj[i]$ with $s[j] = 1$}
                \State $\text{UNION}(UF_{TB},\, i,\, j)$
                \State $\text{UNION}(UF_{LR},\, i,\, j)$
            \EndFor
            \State $C_{TB} \leftarrow \text{CONNECTED}(UF_{TB},\, v_{top},\, v_{bot})$
            \State $C_{LR} \leftarrow \text{CONNECTED}(UF_{LR},\, v_{left},\, v_{right})$
            \If{$\mathcal{M} = \text{Intersection}$ \textbf{and} $C_{TB} \land C_{LR}$} \textbf{break}
            \ElsIf{$\mathcal{M} = \text{Union}$ \textbf{and} $C_{TB} \lor C_{LR}$} \textbf{break}
            \EndIf
        \EndFor
        \State Append $N_{open} / N$ to $\mathbf{P}$
    \EndFor
    \State $\bar{p}_c \leftarrow \frac{1}{T}\sum \mathbf{P}$, \quad
           $\sigma \leftarrow \sqrt{\frac{\sum(\mathbf{P}_i - \bar{p}_c)^2}{T-1}}$, \quad
           $CI_{95} \leftarrow \bigl[\bar{p}_c \pm \tfrac{1.96\,\sigma}{\sqrt{T}}\bigr]$
    \State \Return $\bar{p}_c,\ \sigma,\ CI_{95}$
\end{algorithmic}
\end{algorithm}

\begin{algorithm}[H]
\caption{Monte Carlo Simulation for Smith Hat tile Bond Percolation (Intersection and Union)}
\label{alg:hat_percolation_bond}
\begin{algorithmic}[1]
    \State \textbf{Input:} Set of nodes $V$ (size $N$), Edge list $E$ (size $M$),
    \Statex \hspace{4.2em} Boundary sets $S_{Top}, S_{Bot}, S_{Left}, S_{Right}$
    \State \textbf{Input:} Number of trials $T$, Percolation type $\mathcal{M} \in \{\text{Intersection}, \text{Union}\}$
    \State \textbf{Output:} Mean critical threshold $\bar{p}_c$, Standard deviation $\sigma$, 95\% CI
    \State Initialise results list $\mathbf{P} \leftarrow \emptyset$
    \For{$k = 1$ to $T$}
        \State Initialise $UF_{TB}$ with virtual nodes $v_{top}, v_{bot}$
        \State Initialise $UF_{LR}$ with virtual nodes $v_{left}, v_{right}$
        \Comment{Pre-connect all boundary nodes to virtual nodes at initialisation.}
        \Statex \hspace{4.2em} \Comment{An isolated boundary node cannot bridge to the opposite boundary,}
        \Statex \hspace{4.2em} \Comment{so the bias on $M_{open}/M$ is negligible and vanishes as $L \to \infty$.}
        \For{each $n \in S_{Top}$} $\text{UNION}(UF_{TB},\, v_{top},\, n)$ \EndFor
        \For{each $n \in S_{Bot}$} $\text{UNION}(UF_{TB},\, v_{bot},\, n)$ \EndFor
        \For{each $n \in S_{Left}$} $\text{UNION}(UF_{LR},\, v_{left},\, n)$ \EndFor
        \For{each $n \in S_{Right}$} $\text{UNION}(UF_{LR},\, v_{right},\, n)$ \EndFor
        \State $M_{open} \leftarrow 0$
        \State Generate a uniformly random permutation $\pi$ of $\{0, \dots, M-1\}$
        \For{each bond index $e$ in order $\pi$}
            \State Let $(u, v) \leftarrow E[e]$
            \State $\text{UNION}(UF_{TB},\, u,\, v)$, \quad $\text{UNION}(UF_{LR},\, u,\, v)$
            \State $M_{open} \leftarrow M_{open} + 1$
            \State $C_{TB} \leftarrow \text{CONNECTED}(UF_{TB},\, v_{top},\, v_{bot})$
            \State $C_{LR} \leftarrow \text{CONNECTED}(UF_{LR},\, v_{left},\, v_{right})$
            \If{$\mathcal{M} = \text{Intersection}$ \textbf{and} $C_{TB} \land C_{LR}$} \textbf{break}
            \ElsIf{$\mathcal{M} = \text{Union}$ \textbf{and} $C_{TB} \lor C_{LR}$} \textbf{break}
            \EndIf
        \EndFor
        \State Append $M_{open} / M$ to $\mathbf{P}$
    \EndFor
    \State $\bar{p}_c \leftarrow \frac{1}{T}\sum \mathbf{P}$, \quad
           $\sigma \leftarrow \sqrt{\frac{\sum(\mathbf{P}_i - \bar{p}_c)^2}{T-1}}$, \quad
           $CI_{95} \leftarrow \bigl[\bar{p}_c \pm \tfrac{1.96\,\sigma}{\sqrt{T}}\bigr]$
    \State \Return $\bar{p}_c,\ \sigma,\ CI_{95}$
\end{algorithmic}
\end{algorithm}

The codebase is organised as follows:

\begin{figure}[H]
    \centering
    \includegraphics[width=\textwidth]{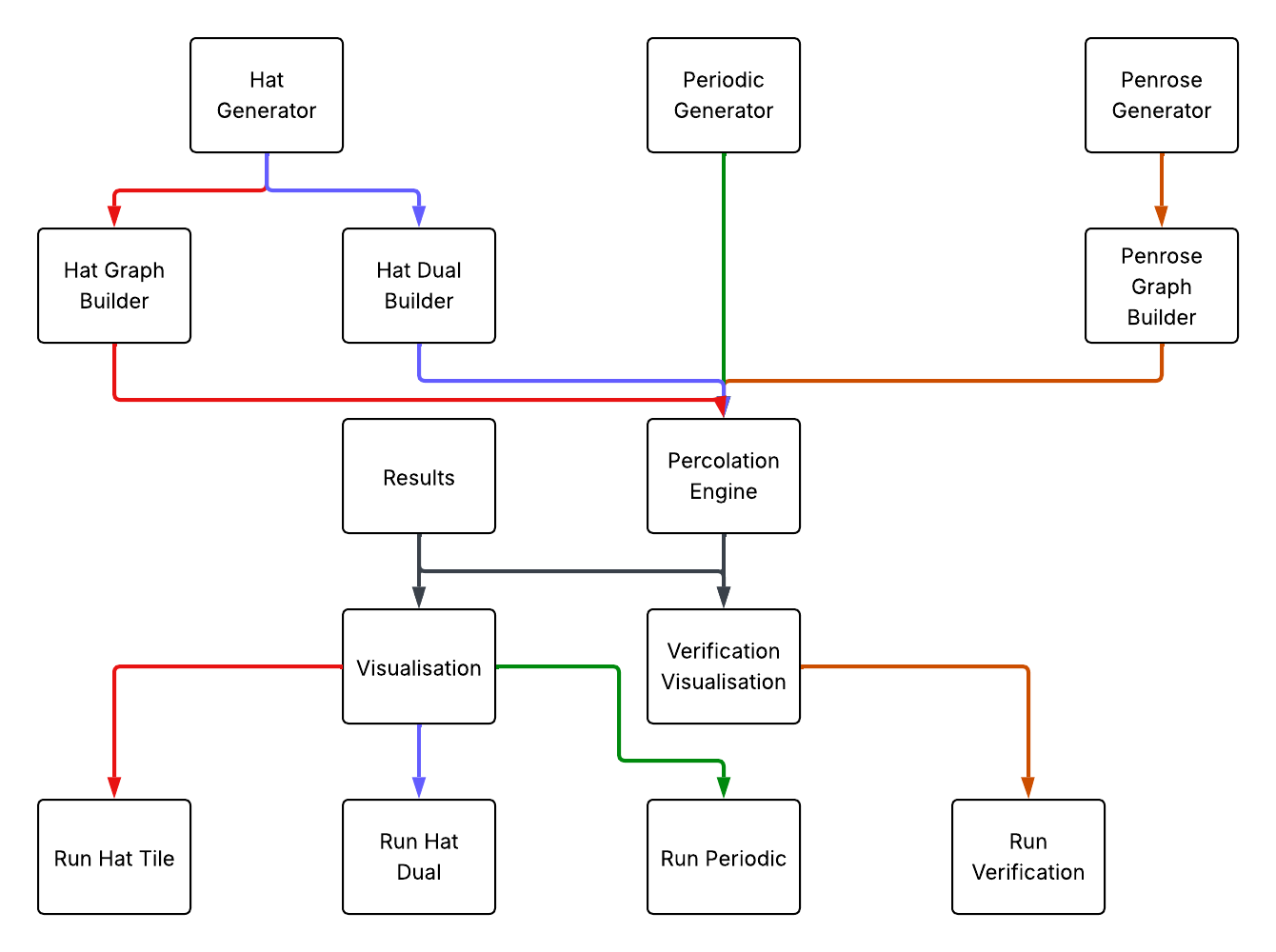}
    \caption{Inter-dependencies of programs within the codebase of our algorithm}
\end{figure}

\pagebreak
\subsection*{Verifications}
The algorithm and pipeline are verified against three geometries with known analytical thresholds: the square lattice, the triangular lattice, and the Penrose P3 tiling — a well-studied aperiodic tiling.

\textbf{Square Lattice.} The square lattice is among the most extensively studied periodic tilings in both theoretical and computational percolation research. Our simulation yields a site percolation threshold of $\hat{p}_c^s = 0.592746$ with 95\% CI $[0.592549, 0.593266]$, in agreement with the high-precision estimate $p_c^s = 0.592746$ reported by \citet{yang2024comment} and \citet{jacobsen2024reply}. For bond percolation, we obtain $\hat{p}_c^b =0.500000$ with 95\% CI $[0.499582, 0.500160]$, consistent with the exact value $p_c^b = 1/2$, which follows rigorously from the self-duality of the square lattice.

\textbf{Triangular Lattice.} The triangular lattice is another canonical periodic structure with known exact thresholds. Our estimate of the site percolation threshold is $\hat{p}_c^s = 0.500000$ with 95\% CI $[0.499386, 0.500142]$, consistent with the exact value $p_c^s = 1/2$ established by self-duality. For bond percolation, we obtain $\hat{p}_c^b = 0.347296$ with 95\% CI $[0.347235, 0.347719]$, in agreement with the exact result $p_c^b = 2\sin(\pi/18) \approx 0.347296$ \citep{sykes1964exact}.

\textbf{Penrose (P3) Tiling.} For site percolation on the P3 Penrose tiling, we obtain $\hat{p}_c^s = 0.584042$ with 95\% CI $[0.582189, 0.585895]$. This is consistent with the high-precision numerical estimate $p_c^s = 0.58391 \pm 0.00001$ of \citet{ZiffBabalievski1999PenroseRhomb} and the earlier result $p_c^s = 0.5842$ of \citet{Sakamoto1989PercolationThresholds}. For bond percolation, we obtain $\hat{p}_c^b = 0.476750 $ with 95\% CI $[0.475409, 0.478090]$, in close agreement with the best available estimate $p_c^b = 0.477$ \citep{yonezawa1988percolation}.




These results confirm that our algorithm and pipeline reproduce known thresholds to high precision, validating the methodology applied to the Smith Hat tile.

\pagebreak
\bibliographystyle{CUP}
\bibliography{bibliography}
\end{document}